\pdfoutput=1
\documentclass[13pt,a4paper,twocolumn]{article}
\usepackage[latin1]{inputenc}
\usepackage[english]{babel}
\usepackage{amsmath}
\usepackage{amsfonts}
\usepackage{amssymb}
\usepackage{mathtools}
\usepackage[pdftex]{graphicx} 
\usepackage{graphicx}
\usepackage[left=2cm,right=2cm,top=2cm,bottom=2cm]{geometry}
\usepackage[justification=centering]{caption}
\usepackage{lipsum}
\usepackage{authblk}
\usepackage{float}

\title{Biased roulette wheel: A Quantitative Trading Strategy Approach}

\author{Giancarlo Salirrosas Mart\'inez, CQF}
\affil{Department of Economics, Universidad San Ignacio de Loyola}
\date{}

\begin{document}
\twocolumn[\begin{@twocolumnfalse}
\maketitle
\vspace{-7 mm}
\begin{abstract}

The purpose of this research paper it is to present a new approach in the framework of a biased roulette wheel. It is used the approach of a quantitative trading strategy, commonly used in quantitative finance, in order to assess the profitability of the strategy in the short term. The tools of \textit{backtesting} and \textit{walk-forward optimization} were used to achieve such task. The data has been generated from a real European roulette wheel from an on-line casino based in Riga, Latvia. It has been recorded  10,980 spins\textsuperscript{1} and sent to the computer through a \textit{voice-to-text} software for further numerical analysis in R. It has been observed that the probabilities of occurrence of the numbers at the roulette wheel follows an Ornstein-Uhlenbeck process. Moreover, it is shown that a flat betting system against Kelly Criterion was more profitable in the short term.  \vspace{7mm}

\end{abstract}

\end{@twocolumnfalse}]

\footnotetext[1]{Data is avaible upon request to: giancarlo.salirrosas@usil.pe}

\section{Introduction}

Roulette wheel is one of the most famous games in the context of casino gambling, although its invention has no clear author, it is meant to be introduced by the french mathematician Blaise Pascal, in his efforts for building a perpetual motion machine around the 17th century . Currently, two types of roulette wheels can be found in a casino: the European and the American. The difference between them relies in the quantity of numbered pockets. An European roulette wheel presents 37 numbered pockets, from 0 to to 36, whereas an American roulette has one additional compartment: the double zero. In both cases, roulette's non-zero numbers are equally distributed into black and red co\-lors, i.e. 18 red and 18 black. Thus, it is clear to realize the house advantage at the American roulette wheel is higher, such that if a player wagers one monetary unit, the average expected value is $-5.3\%$. Meanwhile, in the European roulette the house edge is only $ -2.7\%$ and for that reason it will be matter of study. 

In the game a ball is spun in the opposite direction of a rotating wheel and the profits or losses are in function on whether the ball landed into one of the compartments where the player wagered. The casino offers several options in which gamblers can stablish their bets at the roulette table, each type of them has been assigned a different payout. These stakes can be grouped into \textit{inside bets} and \textit{outside bets}, where the highest payout a player could get belongs to the first group, when a bet is done on a single number - \textit{straight up bet} - the player receives a 35:1 payout. Forecast accurately the next number of the roulette wheel it is considered as the holy grail in the gambling community. 

Since the invention of this game, several approaches and methods were introduced to try to beat roulette, such as martingale systems [1], chaos theory [2], physics [3], and statistical analysis [4],[5],[6]. While the core of martingale systems indicates the gambler to double his bet in a consecutive manner after every loss until a profit occurs which the following form: $2^{0} \cdot x, 2^{1} \cdot x, 2^{2} \cdot x, ..., 2^{n-1} \cdot x$ thus, the cumulative loss is $\sum_{i=1}^{n}(2^{i-1})(x)$, where $x$ is the amount of the initial bet and $n$ the number of bet the gambler has done. This strategy could not be implemented in a real casino due to constraints on the gambler's initial capital - which is finite - and the casino's maximum size per bet limit. Moreover, one of the most recent studies in the field of physics done by Small and Tse (2012),  where they provided a model for the motion of a roulette wheel and ball, showed that knowledge of initial position, velocity and acceleration was sufficient to predict the outcome of roulette and obtain an average expected return of 18\%. 
In the statistical analysis' approach, large samples of data sets are gathered from roulette spins in order to analyse the outcomes and spot patterns to gain an edge over the house. This last approach takes the assumption that a physical imperfection on the roulette wheel may cause deterministic results that players can detect and exploit. Those non random outcomes are translated into a higher than usual probabilities of occurrence, this last approach is also known as the \textit{biased roulette wheel} and will be the focus of this research paper. \\
The contribution of this work it is not only to present a new glance and more robust method to quantify the risk inherent to a biased roulette wheel strategy with tools commonly used in the financial industry, such as portfolio and risk management, but also to assess if this strategy proves to be profitable in the long term. Such task will be achieve through \textit{in-sample} and \textit{out of sample} backtesting, and walk-forward optimization, as if it were a quantitative trading strategy. The lack of an up-to-date biased roulette wheel approach, with a representative sample that accounts for thousands of spins rather than hundreds, and the ambition to beat the game of roulette, inspired this paper.

\section{Data and Methodology}
\subsection{Data}

This work was done with data gathered from a real on-line casino based in Riga, Latvia. The time taken to achieve this task was 23 days (1 July 2016 until 23 July 2016). A collection of 10,980 observations were stored and distributed into 5,000 spins for \textit{in sample} analysis and 5,980 spins for \textit{out of sample} statistical analysis. In order to test the strategy into different time horizons, the \textit{out of sample} group has been divided into seven different subgroups. Each subgroup division has different time intervals.

The reason to collect a large sample size it is because it allows to have more accurate information about the population, as it reflects more reliably the population mean and other statistical properties that are discussed in the theoretical framework section. While larger the sample size, better level of confidence in the sample estimates, and lower uncertainty about the matter of study.

Since the data was not provided by the casino, nor was requested, this has been taken through screen-shots that allowed to visualize the last 500 outcomes from the roulette's spins. First, the raw information was stored as an image data type, then it was reproduced to the computer through a \textit{voice-to-text} software, in order to avoid typographical errors. Finally, this text file was converted into a numerical data type with an algorithm in R. 

\subsection{Methodology}
\subsubsection{Theoretical Framework}

Let $X = \{x_{1}, x_{2}, x_{3}, ..., x_{n}\}$ be a finite set that represents a discrete random variable with a population of size $n$, where $x_{i}$ symbolize the elements of $X$. Let $\Omega$ be  a set of all possible outcomes of $X$, the \textit{sample space}. Finally, let $R$ be a $\sigma$-algebra of subsets of the sample space $\Omega$, i.e a collection of one or more outcomes. Such that, if we spin the roulette wheel $n$ times, the measurable space components are:

\begin{equation}
 \Omega_{n} = \{ X: X = (x_{1}, ..., x_{n}), x_{i}: 0, 1, 2, ..., 36   \} 
  \label{eq:samplespace}
\end{equation}

\begin{equation}
 R = \{ x_{i} \in \Omega_{n} \}
 \label{eq:S} 
\end{equation}

Where $\{x_{i}\}$ is an elementary event. The probability measure $\mathbb{P}$ defined over the sample space $\Omega$, such that the probability of an event $R \subset \Omega $ is denoted as $\mathbb{P}(R) = \sum_{x \in R} p(x)  $. Thus, the probability space is defined as ($\Omega, R, \mathbb{P}$), which satisfies the following axioms:

\begin{equation}
 \forall x \in R : 0\leq \mathbb{P}(x)\leq 1
 \label{eq:ax1}
\end{equation}

\begin{equation}
 \ \mathbb{P}(\Omega) = 1 
  \label{eq:ax2}
\end{equation}

\begin{equation}
 \ \mathbb{P}(\bigcup_{i=1}^{n} R_{i}) =  \sum_{i=1}^{n}\mathbb{P}(R_{i})
  \label{eq:ax3}
\end{equation}

For (\ref{eq:ax3}) we let $R_{1}, ...,R_{n}$ be a countable sequence of pairwise disjoint events. Axioms (\ref{eq:ax1}), (\ref{eq:ax2}) and (\ref{eq:ax3}) were published  initially by Andrei Kolmogorov in his work \textit{Grundbegriffe der Wahrscheinlichkeitsrechnung} in 1956.

Under the classical approach we can define the pro\-bability of an event, and an elementary event as:

\begin{equation}
 \ \mathbb{P}(R) = \dfrac{|R|}{|\Omega|} 
  \label{eq:eq6}
\end{equation}

\begin{equation}
 \ \mathbb{P}(\{x\}) = \dfrac{1}{|\Omega|} = \dfrac{1}{n^{r}}  
  \label{eq:eq7}
\end{equation}

Where in (\ref{eq:eq6}), we note $|R|$ and $|\Omega|$, as the cardinals of $R$ and $\Omega$, i.e. the number of elements in the event subset and the sample space set respectively. In (\ref{eq:eq7}), we note $r$ as the number of trials, and we make the assumption that each elementary event, $x$, has the same probability of occurrence. \\

\noindent \textbf{Example 2.2.1.} A ball is spun once in an unbiased  roulette wheel, i.e a roulette wheel where each outcome has the same probability of occurrence, such that $ \mathbb{P}(x = 0) =  \mathbb{P}(x = 1) = \mathbb{P}(x = 2) = ... = \mathbb{P}(x = 36)$. The probability space $ (\Omega, R, \mathbb{P})$ is defined as: \\

\begin{center}
$\Omega = \{0, 1, 2, ..., 36 \}$ \\
\end{center}

\begin{center}
$R = \{x_{i}\}$ \\
\end{center}

\begin{center}
$ \mathbb{P}(R) = \dfrac{1}{37} \approx 0.027  $ \\
\end{center}

Thus, in this context, it is convenient to state the concept of  probability mass function (PMF), where we note $ F_{X}(x_{i}) = \mathbb{P}(X = x_{i})$, and it is defined as $F_{X}: \mathbb{R} \rightarrow [0, 1] $. The first property of the mass function states that $F_{X}(x_{i}) \geq 0, \forall \ x_{i} $ in the state space, and the second property is $ \sum_{x}F_{X}(x_{i}) = 1 $. In this example, our PMF is defined as an uniform distribution:\\

\begin{equation}
  F_{X}(x_{i})=\begin{cases}
    \dfrac{1}{37}, & \text{if $x \in \{0, 1, 2, ..., 36\}$},\\
    \ 0, & \text{if $x \notin \{0, 1, 2, ..., 36\}$}.
  \end{cases}
  \label{eq:eq08}
\end{equation}

\begin{figure}[H]
  \centering
  \captionsetup{justification=centering}
  \includegraphics[width=3.15in]{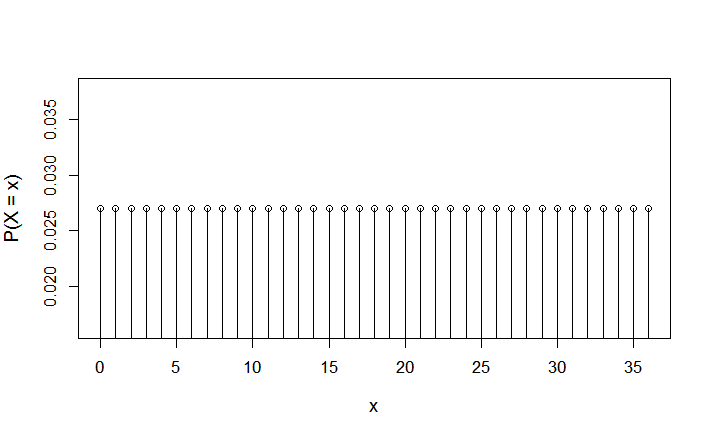}
    \caption{PMF of an unbiased roulette wheel}
   \label{fig: 1}
\end{figure}

Moreover, from a \textit{frequentist} approach, the probabi\-lity of occurrence of an event, $x_{i}$, of a discrete random variable $X$ can be measured as the limit of its relative frequency of occurrence in an infinite sequence of repetitions, such that: 

\begin{equation}
 \ \mathbb{P}(X = x_{i}) = \lim_{n \to \infty}\frac{n(x_{i})}{n}
  \label{eq:eq9}
\end{equation}

Where $ n(x_{i}) $ is the number of times the event $x_{i}$ has taken place in the experiment and $n$ is the total number of repetitions. The element $n \to \infty$ satisfies the condition of the Law of Large Numbers, which holds that as the number of observations increases to infinity, the cumulative relative frequency of an event will converge to the \textit{true} probability of the event itself, and it was proved by Etemadi [7] in his work \textit{An elementary proof of the strong law of large numbers} published in 1981. We will review the two types of Law of Large Numbers: the weak law and the strong law.  \\

\noindent \textbf{Theorem 2.2.1} \textit{[Weak Law of Large Numbers]}. Let $ (X_{n})_{n \in \mathbb{N}} $  be a sequence of $i.i.d.$ random variables, each with finite mean $ \mu = \mathbb{E}[X_{i}] $ and variance $ \sigma^{2} = Var[X_{i}]$. We define $ S_{n} = \dfrac{1}{n} \sum_{i = 1}^{n} X_{i} $. Then, $ \forall \ \xi > 0:$ 

\begin{equation}
 \ \lim_{n \to \infty} \mathbb{P}(\mid S_{n} - \mu \mid \geq \xi  ) = 0
  \label{eq:eq10}
\end{equation}

\noindent \textbf{\textit{Proof}}. From \textit{Chebyshev's inequality} we know that:

\begin{equation}
 \ \mathbb{P}(\mid X - \mu \mid \geq \zeta  ) \leq \dfrac{Var[X]}{\zeta^{2}}
  \label{eq:eq11}
\end{equation}

Where $X$ is a discrete random variable. This can be derived straightforward from \textit{Markov's Inequality}, which holds that $ \mathbb{P}(X \geq \zeta) \leq \mathbb{E}[X]/\zeta $, $ \forall \ \zeta > 0 $. \\

We proceed to solve $ Var[X]$ for $ X = S_{n}$: \\

$ Var[S_{n}] = Var \left[\dfrac{X_{1} + X_{2} + X_{3} + ... + X_{n} }{n}\right]     $ \\
\indent $ Var[S_{n}] = \dfrac{1}{n^{2}}Var[X_{1} + X_{2} + X_{3} + ... + X_{n}]$ \\
\indent $ Var[S_{n}] = \dfrac{n\sigma^{2}}{n^{2}} = \dfrac{\sigma^{2}}{n}$ \\

We replace this result in (\ref{eq:eq11}):

\begin{center}
 $ \mathbb{P}(\mid S_{n} - \mu \mid \geq \zeta  ) \leq \dfrac{\sigma^{2}}{n\zeta^{2}} \to 0 $ as $ n \to \infty $  
\end{center}

\noindent  \textbf{Theorem 2.2.2} \textit{[Strong Law of Large Numbers]}. As above, let $ (X_{n})_{n \in \mathbb{N}} $ denote a sequence of independent and identically distributed random variables with common distribution, each having a finite mean $ \mu = E[X] $, and assume $ E(X_{i}^{4}) < \infty.$ Then :

\begin{equation}
\ \ \ \ \mathbb{P}\left(\lim_{n \to \infty}  \dfrac{X_{1} + X_{2} + ...+ X_{n}}{n} - \mu = 0 \right) = 1 
  \label{eq:eq12}
\end{equation}

\noindent \textbf{\textit{Proof}}. From \textit{Borel-Cantelli} we know that $ if \sum_{n} \mathbb{P}(A_{n})< \infty $, then: 

\begin{equation}
\ \mathbb{P}[A_{n} \ i.o] = 0 
	\label{eq:eq13}
\end{equation}

Where in (\ref{eq:eq13}), $A_{n}$ is a sequence of events. The proof of the proposition from above is straight forward:

\begin{center}

$ \{A_{g} \ i.o.\} \subseteq \cup_{m \geq n} A_{n} $, for any $n$, thus: 

\end{center}

\begin{center}

$\mathbb{P}[A_{g} \ i.o.] \leq \mathbb{P}[\cup_{m \geq n}A_{m}] $
\end{center}

\begin{center}

$\mathbb{P}[A_{g} \ i.o.] \leq \sum_{m \geq n}\mathbb{P}[A_{m}] $ 
\end{center}

If we let $ n \to \infty $, then $ \mathbb{P}[A_{g} \ i.o.] = 0 $, which concludes de proof of the proposition. \\

Let us define,  $A_{n}^{\zeta}$, such that:

\begin{equation}
A_{n}^{\zeta} = \bigg\lbrace X \in \mathbb{R}^{\infty}: \left| \dfrac{1}{n} \sum_{i=1}^{n}X_{i} - \mu \right| \geq \zeta \bigg\rbrace, \ \forall \ \zeta > 0
\label{eq:eq14}
\end{equation}

Where in (\ref{eq:eq14}), $\zeta$ is fixed. We define $Y_{i}\colon = X_{i} - E[X]$ with $ E[Y_{i}] = 0 $, and a finite four moment $E[Y_{i}^{4}] < \infty $. Then:

\begin{center}
$ \dfrac{Y_{1} + Y_{2} + \dots + Y_{n}}{n} \to 0, $ \textit{almost surely.}
\end{center}

From the fourth moment of \textit{Markov's Inequality}, we apply:

\begin{equation*}
 \mathbb{P}(A_{n}^{\zeta}) = \mathbb{P}\left( \left| \dfrac{\sum_{i=1}^{n}Y_{i}}{n} \right| \geq \zeta \right)  \leq \dfrac{1}{\zeta^{4}}  E\left[\left(\dfrac{\sum_{i=1}^{n}Y_{i}}{n}\right)^{4}\right]
 \end{equation*}

We proceed to solve with \textit{Lyapounov's inequality} which holds that $E^{1/\phi}[\mid X \mid ^{\phi}] \leq E^{1/\rho}[\mid X \mid ^{\rho}]$ for $0<\phi<\rho$:

\begin{multline*}
\mathbb{P}(A_{n}^{\zeta}) = \dfrac{1}{n^{4} \zeta^{4}} \left( \sum_{i=1}^{n}E[Y_{i}^{4}] + {4\choose 2} \sum_{i=1}^{n} \sum_{j=i+1}^{n}E[Y_{i}^{2}]E[Y_{j}^{2}] \right) \\ \leq \dfrac{1}{n^{4} \zeta^{4}} \left( \sum_{i=1}^{n}1 + {4\choose 2} \sum_{i=1}^{n} \sum_{j=i+1}^{n}1 \right) E[Y_{i}^{4}]
\end{multline*}

This can be simplified to: 

\begin{equation}
 \mathbb{P}(A_{n}^{\zeta}) = \dfrac{(3n - 2)E[Y_{i}^{4}]}{n^{3} \zeta^{4}}
 \label{eq:eq15}
\end{equation}

Where in (\ref{eq:eq15}), we can realize it is a summable expression. Thus:

\begin{equation}
 \sum_{n=1}^{\infty}\mathbb{P}\left[ \left| \dfrac{1}{n} \sum_{i=1}^{n}X_{i} - \mu \right| \geq \zeta \right]< \infty, \ \forall \ \zeta > 0
\end{equation}

The result from above implies that $(\sum_{i=1}^{n}X_{i})^{4}/n^{4}$ converges to 0 with probability equal to $1$, and thus, $ (\sum_{i=1}^{n}X_{i})/n \to 0 $ also converges with probability 1. This completes the proof of the strong law of large numbers. \\

We note our cumulative distribution function (CDF) as $G_{X}(x_{i})$ defined by $G_{X} : \mathbb{R} \rightarrow [0, 1]$, such that: 

\begin{equation}
G_{X}(x_{i}) = \mathbb{P}(X \leq x_{i}) = \sum_{x_{1} \leq x_{2}}g(x_{1}), \ \forall \ x_{i} \in \mathbb{R}
 \label{eq:eq17}
\end{equation}

A cumulative distribution function has the following properties:\\

\noindent a). $G_{X}({x_{i}})$ is a non-decreasing \textit{step} if $x_{1} \leq x_{2}$ then $G_{X}({x_{1}}) \leq G_{X}({x_{2}})$.\\ 
\noindent b). $0 \leq G_{X}(x_{i}) \leq 1, \ \forall \ x_{i} \in \mathbb{R}$. \\
\noindent c). $G_{X}(x_{i}^{+}) = G_{X}(x_{i}) \ \forall \ x_{i} \in \mathbb{R}.$ Thus, $G_{X}$ is $right-continuous$. \\
\noindent d). $G_{X}(x_{i}^{-}) = \mathbb{P}(X \leq x_{i}) \ \forall \ x_{i} \in \mathbb{R}.$ Thus, $G_{X}$ has \textit{limits from the left}. \\
\noindent e). $ \lim_{x_{i} \to \infty} G_{X}(x_{i}) = 1$. \\
\noindent f). $ \lim_{x_{i} \to -\infty} G_{X}(x_{i}) = 0$. \\

If we take the function defined in (\ref{eq:eq08}), we can define its cumulative distribution function as:

\begin{equation}
  G_{X}(x_{i})=\begin{cases}
    \dfrac{1}{37}, & \text{if \ $0 \leq x < 1$},\\
    \\
    \dfrac{2}{37}, & \text{if \ $1 \leq x < 2$},\\
    \\
    \dfrac{3}{37}, & \text{if \ $2 \leq x < 3$}, \\
    \\
    \vdotswithin{\dfrac{37}{37}} & \vdotswithin{\text{if \ $x \geq 36 $}} \\
    \\
    \dfrac{37}{37}, & \text{if \ $x \geq 36 $}.
  \end{cases}
  \label{eq:eq18}
\end{equation}
 
\begin{figure}[H]
  \centering
  \captionsetup{justification=centering}
  \includegraphics[width=3.20in]{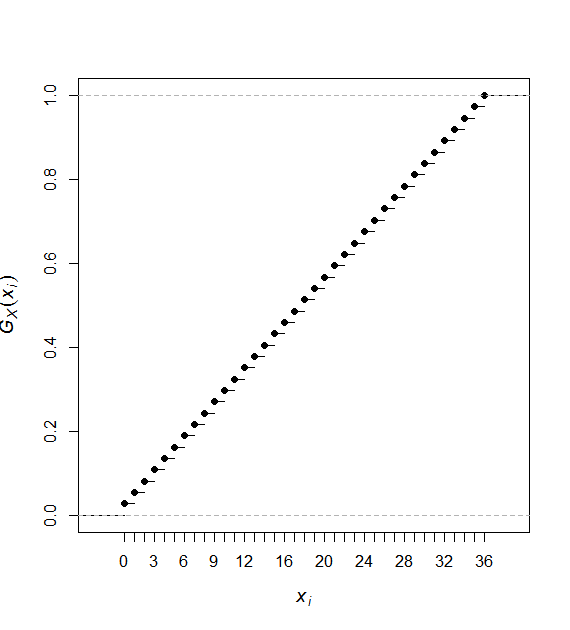}
    \caption{CDF of an unbiased roulette wheel}
   \label{fig: 2}
\end{figure}

In order to assess the fairness of a roulette wheel, we will introduce the concept of \textit{Chi square} test. This statistical test allows to test for deviations of observed frequencies from expected ones:

\begin{equation}
X^{2}_{c} = \sum \dfrac{(O_{i}-E_{i})^{2}}{E_{i}}
 \label{eq:eq19}
\end{equation} 

Where in (\ref{eq:eq19}), $O_{i}$ represents the observed frequencies, and $E_{i}$ symbolizes the expected of the frequencies. If the $X^{2}_{c}$ is less than the critical value of the $Chi square$ distribution associated with $v$ degrees of freedom and a significance level, $\alpha$, we can reject or accept our null hypothesis: \\

$H_{0}$: \textit{The roulette wheel is not biased}. \\
\indent $H_{1}$: \textit{The roulette wheel is biased}.

\subsubsection{Backtesting}

Backtesting is an indispensable tool for decision ma\-king commonly used  in quantitative finance [8],[9],[10], in order to assess the efficiency and viability of a tra\-ding strategy before to be implemented in real market conditions. A backtest it is conduced in order to see if the final results meet the initial criteria, spot the possible drawbacks, and analyse the profit or losses of the strategy itself.  In the financial markets, most of these trading strategies are executed in an automated manner in order to avoid decisions based on human emotions, and in some cases for \textit{speed} purposes, like in the field of High Frequency Trading [11],[12],[13]. 

In this research, we will use backtesting to asses whether the \textit{biased wheel} approach is profitable in the short-term or not, as well to spot the risks of following this method, and analyse other statistical properties of the strategy. In order to define our \textit{roulette strategy}, we treat the concept of expected value for a discrete random variable:

\begin{equation}
E(X) = \sum_{i=1}^{n}x_{i}\mathbb{P}(x_{i})
\label{eq:eq20}
\end{equation}

Where for our strategy, in (\ref{eq:eq20}), $x_{i}$ is the payoff  and $\mathbb{P}(x_{i})$ is the probability of that payoff takes place, such that if we bet $\$1$ on a single number per spin in an unbiased roulette wheel our expected value would be:

\begin{equation}
E(X) = (\$35)(\dfrac{1}{37})+ (-\$1) (\dfrac{36}{37}) \approx -\$0.027
\label{eq:eq21}
\end{equation}

This means, that for each dollar we bet, we loose $ -\$0.027$ or $-2.7\%$. Now, if the roulette wheel is biased, the probability of occurrence of certain numbers could be greater than $1/37$, let us make the assumption of a case where $\mathbb{P}(x_{1}) = 3\%$:

\begin{equation}
E(X) = (\$35)(3\%)+ (-\$1) (97\%) \approx \$0.08
\label{eq:eq22}
\end{equation}

The expectation in (\ref{eq:eq22}) is $\$0.08$ for a dollar, or $8\%$, almost four times higher than the normal expectation of $-\$0.027$ per spin. \\
As show in (\ref{eq:eq22}), a probability of occurrence of $3\%$ can increase the expected value dramatically. We will use the $3\%$ level as a filter for selecting the numbers to bet. From this we begin to  compute the filter of our backtester as boolean, where $1$ means \textit{True} and $0$ stands for \textit{False}, as follows:

\begin{equation}
  B_{X}(x_{i})=\begin{cases}
    1, & \text{if \ $ \mathbb{P}(X = x_{i}) \geq 3.0\% $},\\
    0, & \textit{otherwise}.
  \end{cases}
  \label{eq:eq23}
\end{equation}

We define the payoff function for a single number bet as:

\begin{equation}
J_{X}(x_{i})=\begin{cases}
    35b, & \text{if \ $ B_{X}(x_{i}) = 1$},\\
    -b, & \textit{if \ $ B_{X}(x_{i}) = 0$}.
  \end{cases}
  \label{eq:eq24}
\end{equation}

Where in (\ref{eq:eq24}), $b$ is the amount willing to risk per bet. Thus, the final value of our portfolio is composed such as:

\begin{equation}
W_{X} = \sum_{i=1}^{n}J_{X}(x_{i}) + I_{c}
\label{eq:eq24} 
\end{equation}

Where $W_{X}$ represent the value of our portfolio, and $I_{c}$ denotes the amount of initial capital. \\

One of the most important aspects in \textit{portfolio management} is \textit{money management}, use the optimal amount of capital to obtain the best return possible at a certain level of risk. Otherwise, if we overuse our capital, or take a higher risk than necessary, even if our strategy is robust and proves to be profitable, a couple of consecutive losses are enough to make our portfolio go to zero. This phenomena is also known as the \textit{gambler's ruin}. Thus, it is important to establish a framework which allow us to find the optimal bet size for a wager. This concept is known in finance and in probability theory as the \textit{Kelly criterion}, and was introduced by John Larry Kelly Jr.[14] in 1956.

\begin{equation}
K = \dfrac{Bp-q}{B}
\label{eq:eq26}
\end{equation}

Where $B$ is the payout, $p$ is the probability of the desired event takes place, $q = 1-p$, and $K$ is expressed as the fraction of the initial capital to wager. 
In an unbiased roulette wheel, the Kelly criterion is:

\begin{equation*}
K = \dfrac{(35)\left(\dfrac{1}{37}\right) - \dfrac{36}{37}}{35} \approx -0.077\%
\end{equation*}

As $K < 0 $, according to Kelly criterion, we should not make the bet. Moreover, if we apply our filter of a probability greater or equal than three percent, the result turns to be different:

\begin{equation*}
K = \dfrac{(35)(3\%) - 97\%}{35} \approx 0.229\%
\end{equation*}

Now, for $K=0.229\%$, then we should wager $0.229\%$ of our current initial capital, $I_{c}$, for a single number. \\

Another important aspect to take into account is \textit{risk management}. There exist several approaches that measure risk in a systematic way, such as standard deviation parameters, but for testing how much can handle our initial portfolio to the fluctuations inherent to our quantitative trading strategy, the concept of \textit{maximum drawdown} is more useful. \\

Maximum drawdown, \textit{(MDD)}, is the maximum cumulative loss from a peak to a bottom on a time interval $[0,T]$, i.e. is the largest drop from the highest value of our portfolio to its lowest point. Let us define $M(T)$ as the maximum drawdown of our portfolio, $W_{X}$, at time $T$:

\begin{equation}
M(T)= \sup_{t \in [0,T]}\left[\sup_{s \in [0,t]}W_{X}(s)-W_{X}(t)  \right]
\end{equation}

A high level of maximum drawdown indicates that the strategy is exposed to a higher downside risk, whereas a low level shows that the strategy can have a sustain growth over time at a low level of risk. The maximum drawdown can also be used as a factor of a risk-adjusted measure of performance such as the \textit{Calmar} ratio, $C(T)$, which was introduced by Terry W. Young [15] in 1991:

\begin{equation}
C(T)= \dfrac{\textit{Return over \ } [0,T]}{M(T) \ \textit{over \ } [0,T]}
\end{equation}

This ratio will be used instead of the Sharpe Ratio due to a lack of a comparable risk free rate, to calculate the return of our strategy relative to downside risk. The higher the value of $C(T)$, the better the strategy performed over the time interval $[0,T]$.
 
\subsubsection{Walk-Forward Optimization}

A \textit{walk-forward optimization} is an iterative function where first, the data is segregated into multiple segments, then  the parameters obtained from the in-sample optimization are backtested in the out-of-sample data, in order to avoid \textit{curve fitting} issues. In this paper, it is used an \textit{anchored walk-forward optimization}, due to the preference of a larger in-sample length. In this class of iterative function, the initial point for optimization is the same for both, in-sample and out-sample data, i.e. as this process continues, the out-of-sample data is added to a fixed in-sample data set to test a new out-of-sample data. 

\section{Results}

In this section, it is presented the results of this research paper. First, it is shown the convergence pro\-bability of each number of the roulette wheel $[0, 36]$ for the in-sample data analysis. Then, it is presented the backtesting results, for the in-sample data. Finally, it is applied the walk-forward optimization in order to assess the profitability of the strategy.

\begin{figure}[H]
  \centering
  \captionsetup{justification=centering}
  \includegraphics[width=7.10in, height= 8.60in]{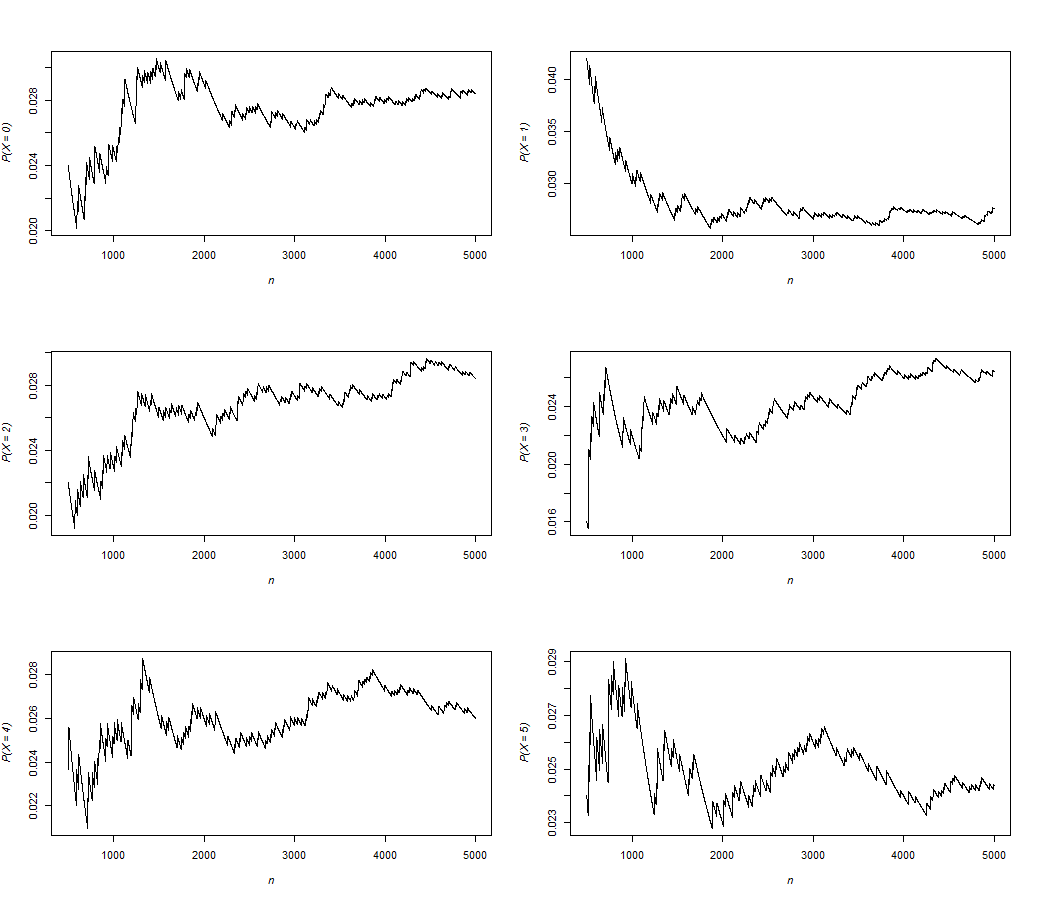}
    \caption*{}
\end{figure}
\clearpage
\begin{figure}[H]
  \centering
  \captionsetup{justification=centering}
  \includegraphics[width=7.10in, height= 8.60in]{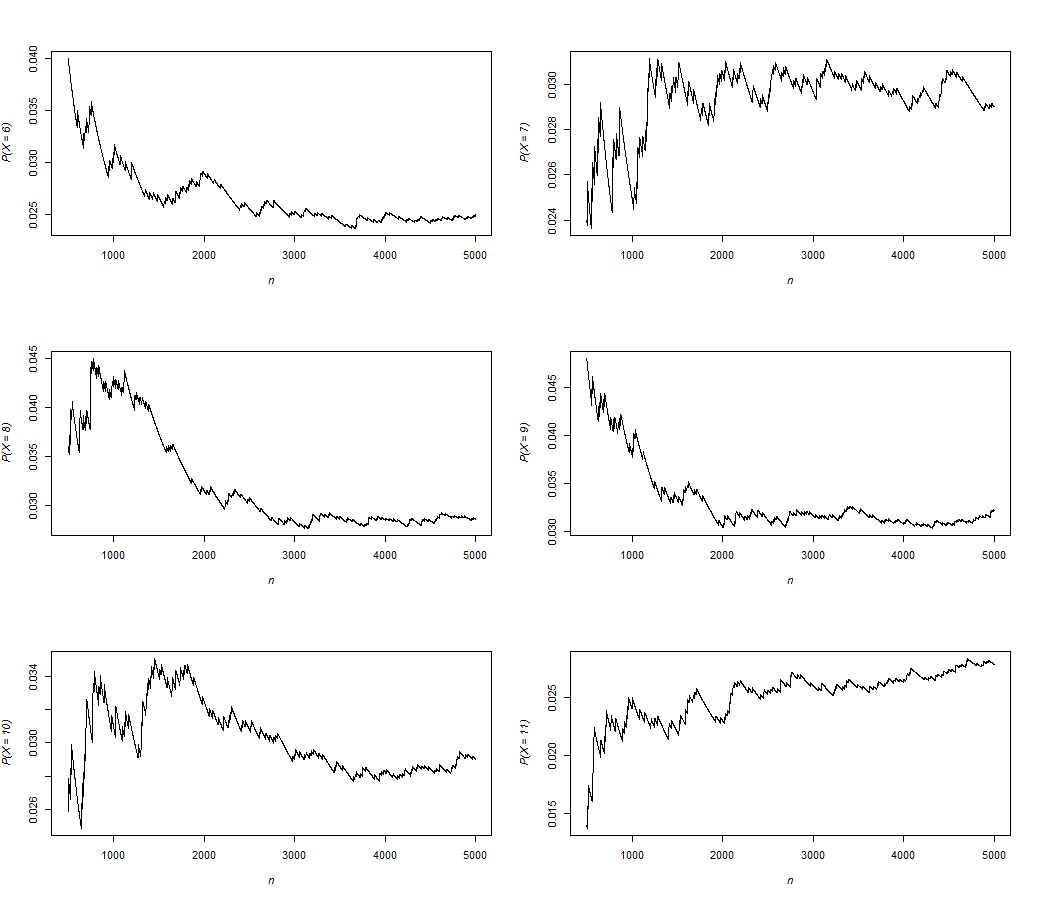}
    \caption*{}
\end{figure}
\clearpage
\begin{figure}[H]
  \centering
  \captionsetup{justification=centering}
  \includegraphics[width=7.10in, height= 8.60in]{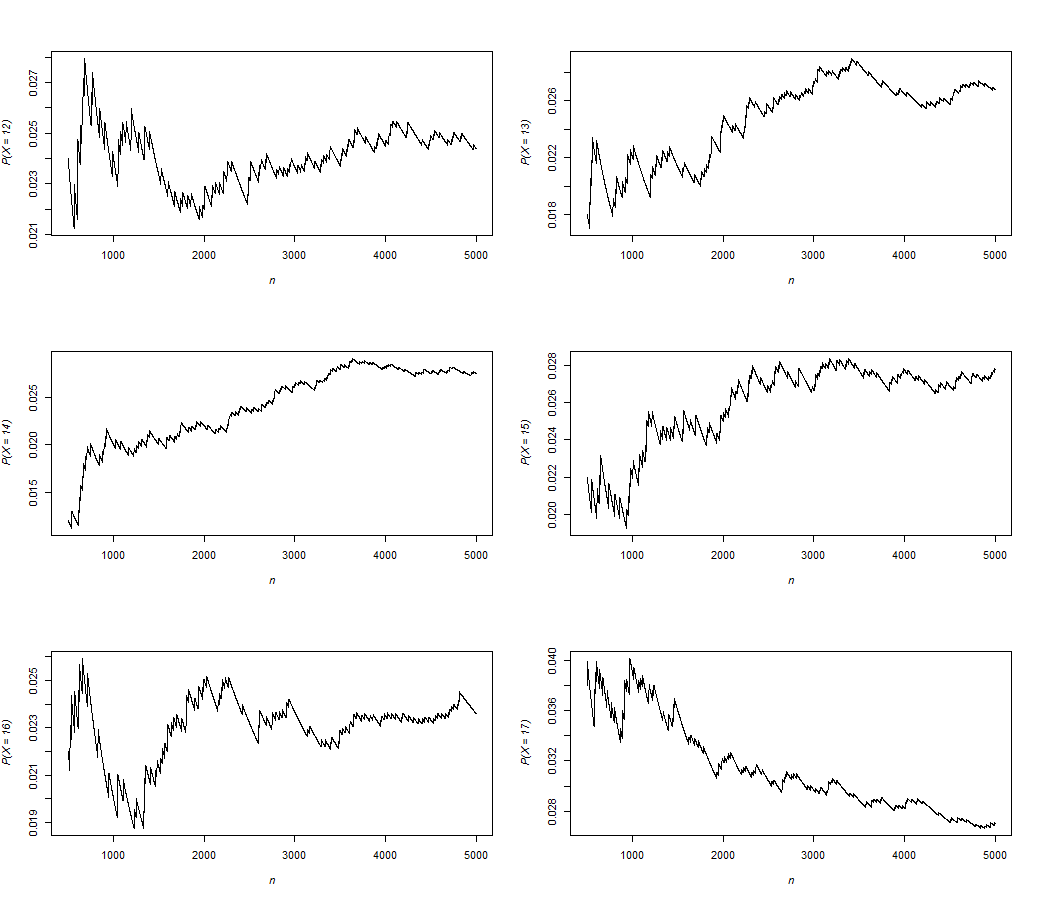}
    \caption*{}
\end{figure}
\clearpage
\begin{figure}[H]
  \centering
  \captionsetup{justification=centering}
  \includegraphics[width=7.10in, height= 8.60in]{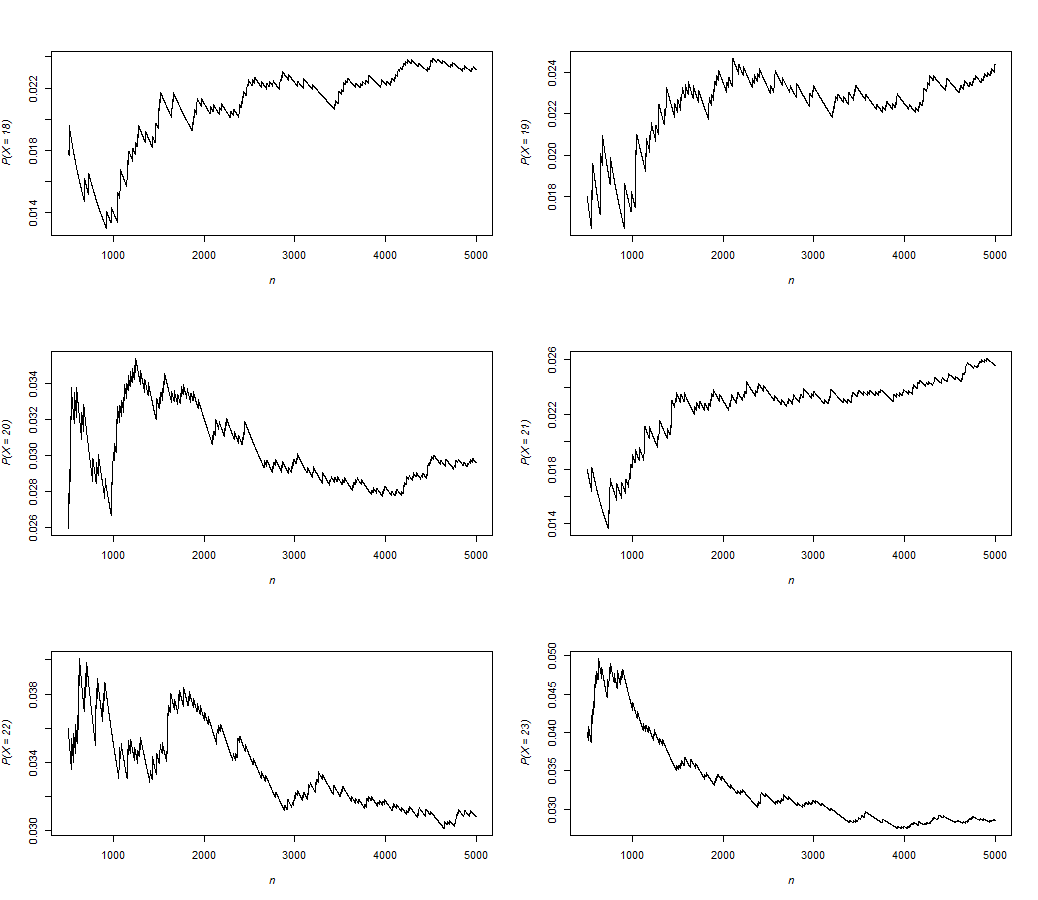}
    \caption*{}
\end{figure}
\clearpage
\begin{figure}[H]
  \centering
  \captionsetup{justification=centering}
  \includegraphics[width=7.10in, height= 8.60in]{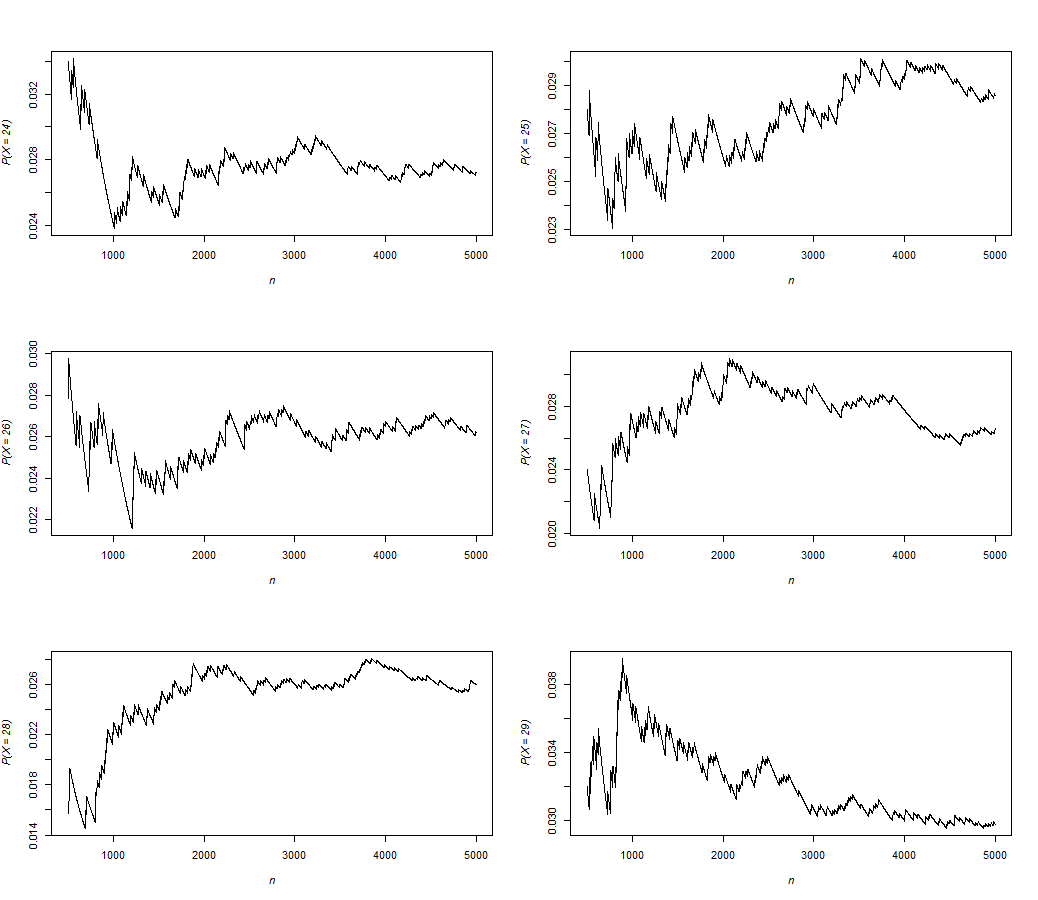}
    \caption*{}
\end{figure}
\clearpage
\begin{figure}[H]
  \centering
  \captionsetup{justification=centering}
  \includegraphics[width=7.10in, height= 8.60in]{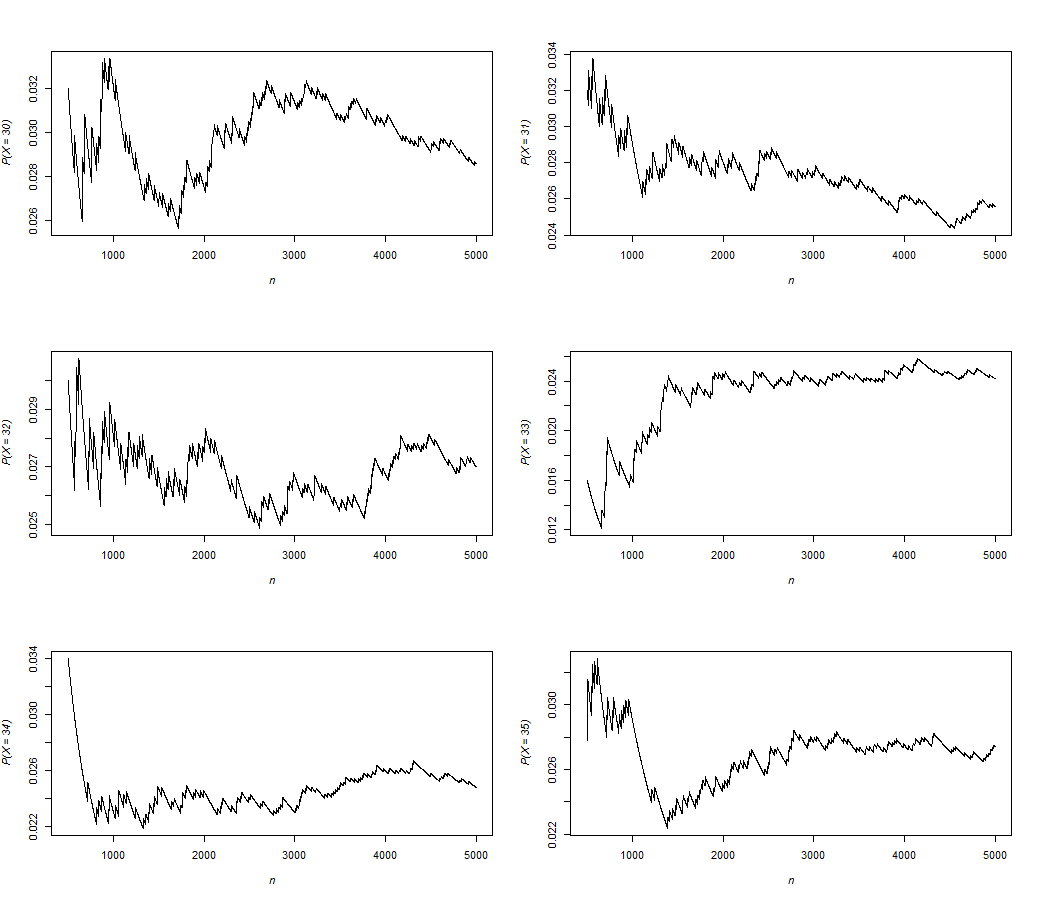}
    \caption*{}
\end{figure}
\clearpage
\begin{figure}[H]
  \centering
  \captionsetup{justification=centering}
  \includegraphics[width=7.10in, height= 8.60in]{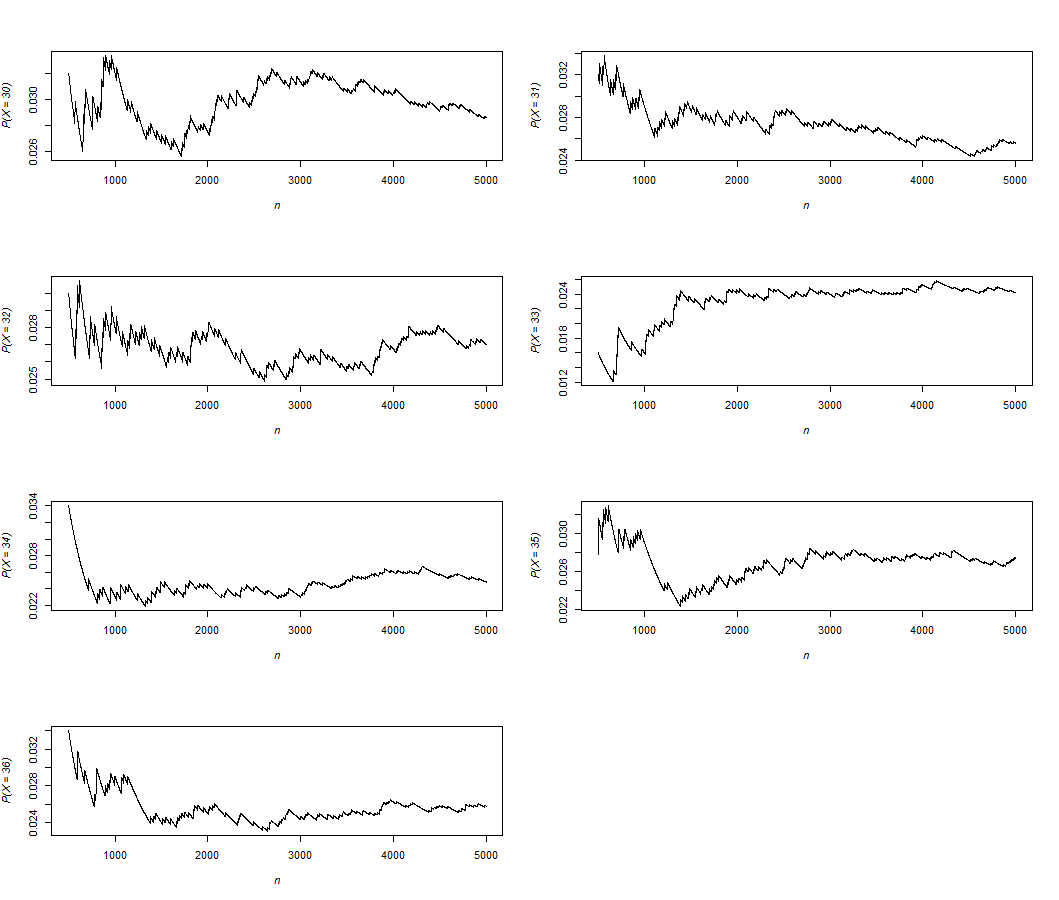}
    \caption*{}
\end{figure}
\clearpage

\begin{table}[H]
\centering
	\begin{tabular}{c c c c}
	\hline 
	\textit{Outcome} & \textit{Probability} & \textit{Mean}& \textit{St. Dev} \\ [0.5ex]
\hline \\
0	& 2.84\%	& 2.74\%	& 0.18\% \\
1   & 2.76\% 	& 2.82\%	& 0.28\% \\
2   & 2.84\% 	& 2.67\%	& 0.21\% \\
3   & 2.64\% 	& 2.43\%	& 0.17\% \\
4	& 2.60\% 	& 2.60\%	& 0.13\% \\
5   & 2.44\% 	& 2.50\%	& 0.11\% \\
6   & 2.50\% 	& 2.67\%	& 0.29\% \\
7   & 2.90\% 	& 2.94\%	& 0.14\% \\
8   & 2.86\% 	& 3.22\%	& 0.50\% \\
9   & 3.22\% 	& 3.33\%	& 0.36\% \\
10  & 2.90\%	& 3.30\%	& 0.21\% \\
11  & 2.78\% 	& 2.51\%	& 0.21\% \\
12  & 2.44\% 	& 2.51\%	& 0.11\% \\
13  & 2.68\% 	& 2.48\%	& 0.29\% \\
14  & 2.76\% 	& 2.41\%	& 0.40\% \\
15  & 2.78\% 	& 2.59\%	& 0.22\% \\
16	& 2.36\% 	& 2.30\%	& 0.13\% \\
17	& 2.70\% 	& 3.14\%	& 0.35\% \\
18	& 2.32\% 	& 2.08\%	& 0.27\% \\
19	& 2.44\% 	& 2.24\%	& 0.17\% \\
20	& 2.96\% 	& 3.03\%	& 0.20\% \\
21	& 2.56\% 	& 2.25\%	& 0.26\% \\
22	& 3.08\% 	& 3.36\%	& 0.24\% \\
23	& 2.86\% 	& 3.31\%	& 0.58\% \\
24	& 2.72\% 	& 2.78\%	& 0.15\% \\
25	& 2.86\% 	& 2.76\%	& 0.16\% \\
26	& 2.62\% 	& 2.58\%	& 0.11\% \\
27	& 2.66\% 	& 2.75\%	& 0.19\% \\
28	& 2.60\% 	& 2.51\%	& 0.29\% \\
29	& 2.98\% 	& 3.21\%	& 0.21\% \\
30	& 2.86\% 	& 2.98\%	& 0.16\% \\
31	& 2.56\% 	& 2.73\%	& 0.17\% \\
32	& 2.70\% 	& 2.68\%	& 0.10\% \\
33	& 2.42\% 	& 2.29\%	& 0.30\% \\
34	& 2.48\% 	& 2.46\%	& 0.15\% \\
35	& 2.74\% 	& 2.69\%	& 0.17\% \\
36	& 2.58\%	& 2.56\%	& 0.16\% 
	\end{tabular}
	\caption{Probability of occurrence for in-sample data}
\end{table}

As we can see, the numbers that has higher probability of occurrence in the sample of 5,000 spins are: 9, 22 and 29; with probabilities of $3.22\%$, $3.08\%$ and $2.98\%$ respectively. \\

It has also been found, that the probability of occurrence of the roulette wheel numbers seems to follow an stochastic process and, in some specific cases, the way they converge to its mean are more stable than other numbers. This is the case for numbers such as: 1, 6, 8, 9, 17 and 23, which coincidently, are the numbers with higher standard deviations than the other ones: $0.28\%$, $0.29\%$, $0.50\%$, $0.36\%$, $0.35\%$ and $0.58\%$ respectively. \\

We can try to model this stable convergence to the mean as a mean-reverting process, with an \textit{Ornstein-Uhlenbeck process}, which is commonly used in quantitative finance to model interest rates [16], [17], [18].

As the number with the highest probability of occurrence is the number $9$ we will use it to try to model the way its probability evolve over time. We define the following stochastic differential equation as:

\begin{equation}
dP(t) = A(P,t)dt + B(P,t)dW(t)
\end{equation}

Where $P(t)$ is the process of $\mathbb{P}(X=9)$ over a time interval $[0,T]$. In the RHS, the term $A(P,t)dt$ is deterministic, and the coefficient of $dt$ is know as the \textit{drift} of the process; the other term $B(P,t)dW(t)$ is random. The coefficient of $dW(t)$ is known as the \textit{di\-ffusion} of the process, where $W(t)$ follows a standard \textit{Wiener process}. This can be written in integral form as:

\begin{equation}
P(t) = P(0) + \int_{0}^{t}A(P, \tau)d \tau + \int_{0}^{t}B(P, \tau)dW(\tau)
\end{equation}

From this, we can begin to state our \textit{Ornstein-Uhlenbeck process} as:

\begin{equation}
dP(t)= -\theta(P(t) - \mu )dt + \sigma dW(t)
\label{eq:eq31}
\end{equation}

Where $\mu \in \mathbb{R}$, $\sigma > 0$ and $\theta > 0$ are parameters. This can be integrated from $0$ to $t$ with the auxiliary function $f(x,t)= x e^{\theta t}$  to:

\begin{equation}
P(t)e^{\theta t} = P(0) + \int_{0}^{t}e^{\theta \tau}\theta \mu d\tau + \int_{0}^{t} \sigma e^{\theta \tau}dW(\tau)
\end{equation}

The expression from above can be simplified as:

\begin{equation*}
P(t) = P(0)e^{- \theta t} + \mu(1 - e^{-\theta t}) + \sigma \int_{0}^{t}e^{- \theta (t-\tau) } dW(\tau)
\end{equation*}

A set of 1,000 Monte Carlo simulations are performed in \textit{Mathematica} for the \textit{Ornstein-Uhlenbeck process} described in (\ref{eq:eq31}), in order to model the probability of occurrence of the number 9 : \\

\begin{figure}[H]
  \centering
  \captionsetup{justification= justified}  
  \includegraphics[width=3.4in]{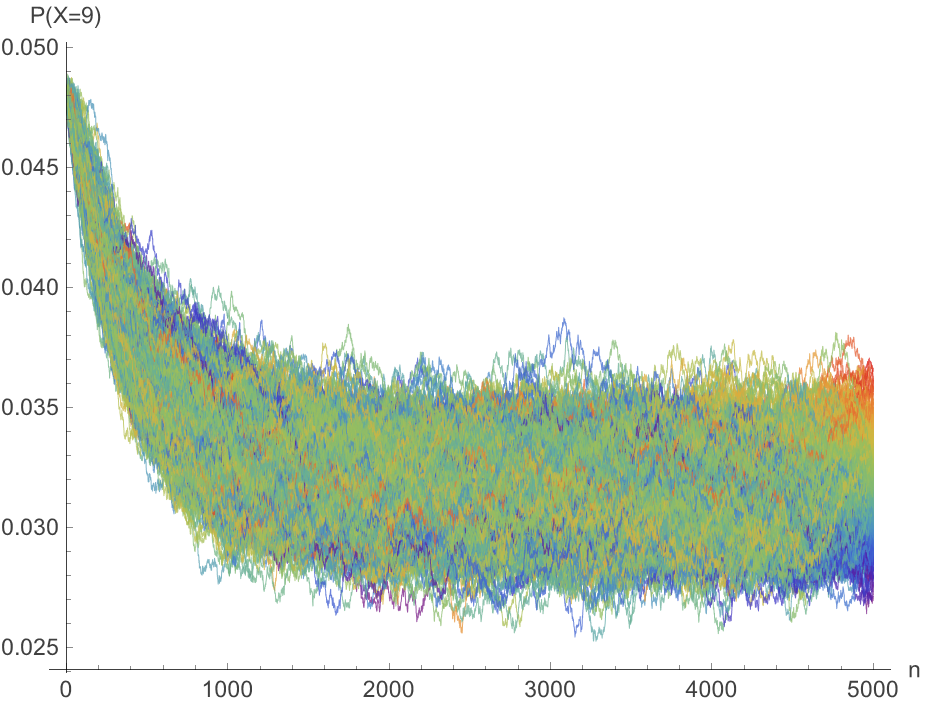}
    \caption{Ornstein-Uhlenbeck process with parameters $\mu = 0.0317$, $\sigma = 0.00010$, $\theta = 0.0022$ and $P(0) = 0.048$}
\end{figure}

\begin{figure*}[ht!]
  \centering
  \captionsetup{justification= justified}  
  \includegraphics[width=\textwidth, height = 11cm]{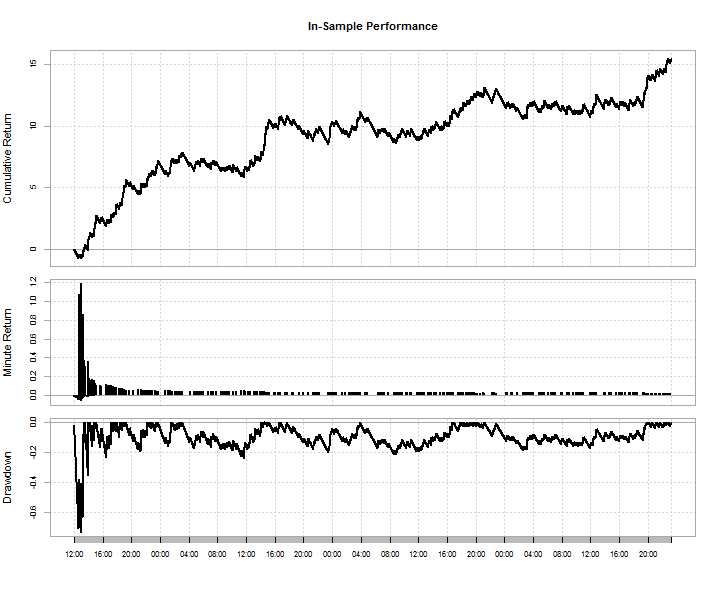}
    \caption{ Cumulative return, returns and drawdown from \textit{in-sample} backtesting for wagers on numbers 9 and 22. Conservative flat betting system established by half a \textit{Kelly Criterion} equal to $1.81 \%$}
\end{figure*}

Despite the fact that the roulette wheel is not statistically biased due to a \textit{Chi square} test, where $ X_{calc}^{2} = 28.1076$ is less than $X_{stat}^{2} = 50.999$ with $36$ \textit{degrees of freedom} and $95\%$ confidence level, we can still have profits if we select only the numbers that pass the filter of greater or equal than three percent of probability of occurrence, which in this case are the numbers 9 and 22. Moreover, we will compare the performance of a flat betting system (FBS) against \textit{Kelly}. The wager's size for each FBS are defined as the average of all Kelly's position size in each Walk-Forward Optimization. An initial capital of $ \$ 2,000 $ was selected arbitrary in order to assess the statistical properties of the strategy. \\

For this \textit{in-sample} backtesting we can see the presence of \textit{curve fitting} which is evident in the distribution of the returns figure from above. This issue is generated because the strategy was designed and over-optimized to this specific data set, i.e. a good metaphor would be: buying a lottery ticket with a previous knowledge of the winning numbers. In order to avoid \textit{curve fitting} issues and test if our strategy it is robust we will use the tool of \textit{walk-forward optimization}. The results from the walk-forward optimization and backtesting are presented in the same way as the back-tester used in \textit{Bloomberg L.P.}'s software as follows:\\

\begin{table}[H]
\centering
\label{my-label}
\caption{Performance Summary}
\begin{tabular}{ll}
\hline 
Number of obs. &  5000 					\\
\\
Wins           & 315                      \\
\\
Losses         & 4685                     \\
\\
P\&L           & 29,979.14                \\
\\
\% P\&L        & 1,498.96                 \\
\\
Total Wins     & 199,710                  \\
\\
Total Losses   & 169,730.9                \\
\\
Avg. Win       & 634                      \\
\\
Avg. Loss      & 36.22                    \\
\\
Max. Win       & 634                      \\
\\
Max. Loss      & 36.22                    \\
\\
Max. Drawdown  & 5,053.89                  \\
\\
Calmar Ratio   & 5.93                  		\\ \hline 
\end{tabular}
\end{table}

\vfill
\vfill
\hfill
\hfill
\vfill
\vfill
\hfill
\hfill

\begin{figure*}[ht!]
  \centering
  \captionsetup{justification= justified}  
  \includegraphics[width=\textwidth, height = 11.5cm]{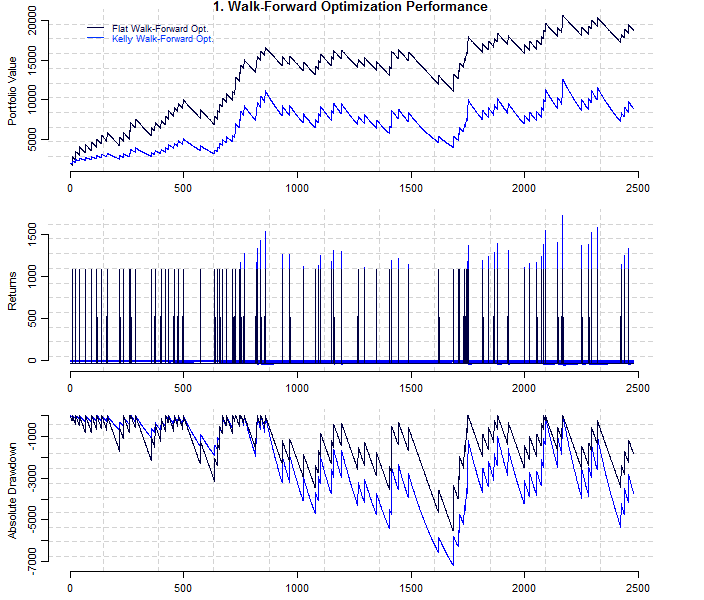}
    \caption{ Wagers only on number 9. Probability of occurrence equal to $ 3.22\%$. FBS' position size is $ \$30.86 $. Kelly's position size is $ 0.4549\% $. Probabilities gathered from in-sample analysis of 5,000 observations.}
\end{figure*}

\begin{table}[H]
\centering
\label{my-label}
\caption{Performance Summary}
\begin{tabular}{c c c}
\hline
\textit{Statistics} & \textit{Kelly Criterion} & \textit{Flat wager}\\ [0.5ex]
\hline \\ 
Number of obs. &  2479 		&	2479			\\
\\
Wins           & 84			&   84                   \\
\\
Losses         & 2395		&   2395                   \\
\\
P\&L           & 6,839.57	&   16,790.21             \\
\\
\% P\&L        & 341.98     &   839.51            \\
\\
Total Wins     & 81,066.99  &   90,741,23                \\
\\
Total Losses   & 74,218.33  &   73,920.15             \\
\\
Avg. Win       & 965.08     &   1080.25                 \\
\\
Avg. Loss      & 30.99      &   30.86              \\
\\
Max. Win       & 1,726.00      &   1080.25                 \\
\\
Max. Loss      & 57.16      &   30.86               \\
\\
Max. Drawdown  & 7,167.49   &   5,524.72              \\
\\
Calmar Ratio   & 0.95       &   3.04            		\\ \hline 
\end{tabular}
\end{table}

\begin{figure}[H]
  \centering
  \captionsetup{justification= justified}
  \includegraphics[width=3.5in, height = 4in]{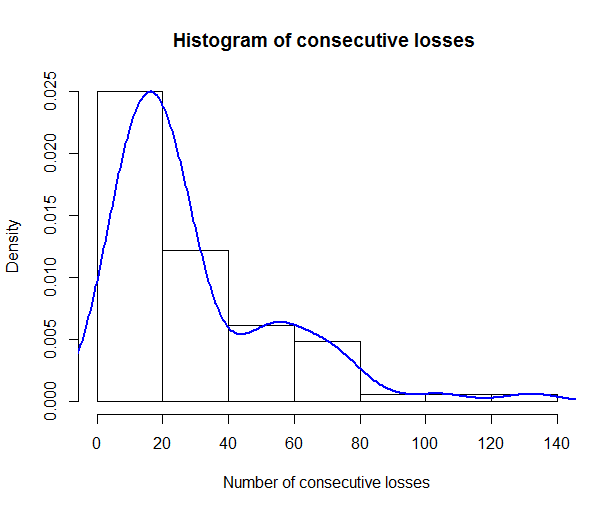}
    \caption{Histogram of consecutive losses for Walk-Forward Optimization No. 1. Maximum consecutive loss equal to 132 losses in a row.}
   \label{fig: 1}
\end{figure}

\begin{figure*}[ht!]
  \centering
  \captionsetup{justification= justified}  
  \includegraphics[width=\textwidth, height = 11cm]{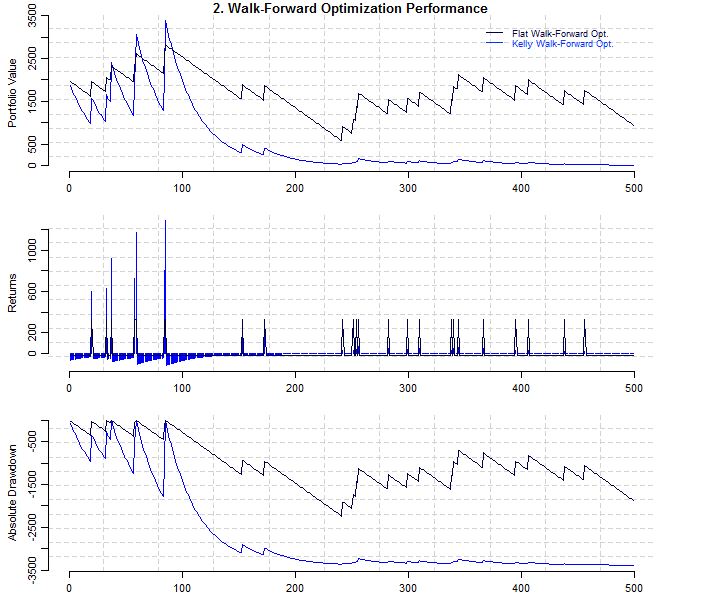}
    \caption{ Wagers on numbers 9 and 20. Probabilities of occurence equal to $ 3.2758\%$ and $ 2.9282\%$ respectively. FBS' position size is $ \$18.81 $. Kelly's position size is $ 3.5242\% $. Probabilities gathered from in-sample analysis of 7,479 observations.}
\end{figure*}

\begin{table}[H]
\centering
\label{my-label}
\caption{Performance Summary}
\begin{tabular}{c c c}
\hline
\textit{Statistics} & \textit{Kelly Criterion} & \textit{Flat wager}\\ [0.5ex]
\hline \\ 
Number of obs. &  499		&	499			\\
\\
Wins           & 24			&   24                   \\
\\
Losses         & 475		&   475                   \\
\\
P\&L           & -1,992.21	&   -1,053.43             \\
\\
\% P\&L        & -99.61     &   -52.67            \\
\\
Total Wins     & 7,012.65  &   7,900.69                \\
\\
Total Losses   & 8,934.38  &   8,935.304             \\
\\
Avg. Win       & 292.19     &   329.20                \\
\\
Avg. Loss      & 18.81      &   18.81              \\
\\
Max. Win       & 1,294.00      &   329.20                 \\
\\
Max. Loss      & 119.60      &   18.81              \\
\\
Max. Drawdown  & 3,384.99   &   2,238.53              \\
\\
Calmar Ratio   & -0.59       &   -0.47            		\\ \hline 
\end{tabular}
\end{table}

\begin{figure}[H]
  \centering
  \captionsetup{justification= justified}
  \includegraphics[width=3.5in, height = 4in]{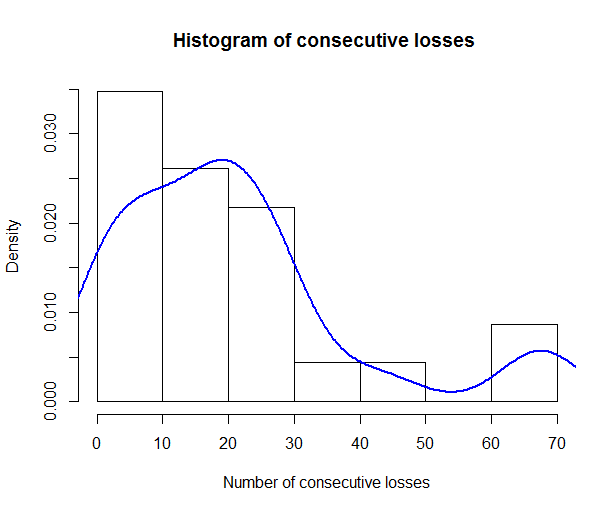}
    \caption{Histogram of consecutive losses for Walk-Forward Optimization No. 2. Maximum consecutive loss equal to 68 losses in a row.}
   \label{fig: 1}
\end{figure}

\begin{figure*}[ht!]
  \centering
  \captionsetup{justification= justified}  
  \includegraphics[width=\textwidth, height = 11cm]{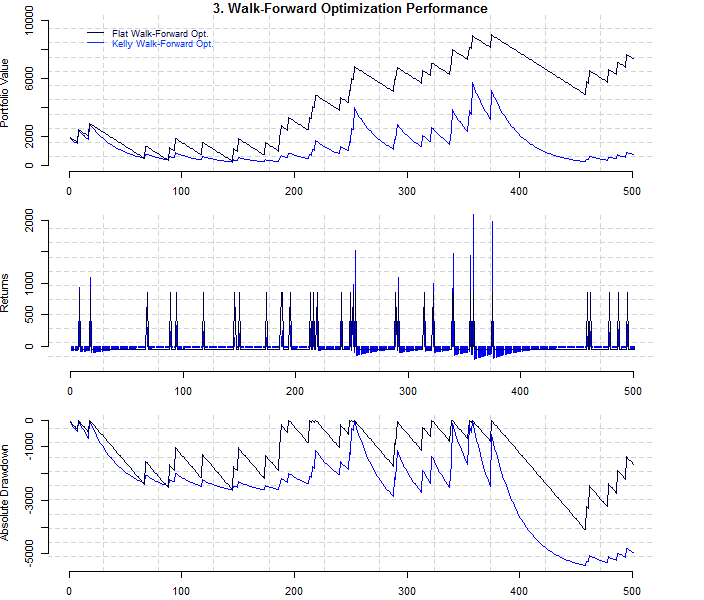}
    \caption{ Wagers on numbers 9 and 10. Probabilities of occurrence equal to $ 3.2214\%$ and $ 3.0083\%$ respectively. FBS' position size is $ \$49.22 $. Kelly's position size is $ 3.5505\% $. Probabilities gathered from in-sample analysis of 7,978 observations.}
\end{figure*}

\begin{table}[H]
\centering
\label{my-label}
\caption{Performance Summary}
\begin{tabular}{c c c}
\hline
\textit{Statistics} & \textit{Kelly Criterion} & \textit{Flat wager}\\ [0.5ex]
\hline \\ 
Number of obs. &  501		&	501			\\
\\
Wins           & 33			&   33                   \\
\\
Losses         & 468		&   468                   \\
\\
P\&L           & -1,269.61	&   5340.02            \\
\\
\% P\&L        & -63.48     &   267.00            \\
\\
Total Wins     & 22,170.33  &   28,422.68                \\
\\
Total Losses   & 23,368.93  &   23,033.45             \\
\\
Avg. Win       & 671.83     &   861.29               \\
\\
Avg. Loss      & 49.93      &   49.22              \\
\\
Max. Win       & 2,181.00      &  861.29                 \\
\\
Max. Loss      & 202.10      &   49.22              \\
\\
Max. Drawdown  & 5,433.79   &   4,084.99              \\
\\
Calmar Ratio   & -0.23       &   1.31            		\\ \hline 
\end{tabular}
\end{table}

\begin{figure}[H]
  \centering
  \captionsetup{justification= justified}
  \includegraphics[width=3.5in, height = 4in]{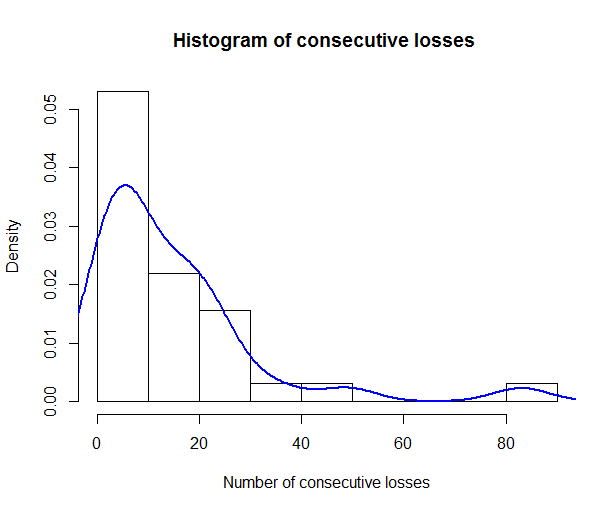}
    \caption{Histogram of consecutive losses for Walk-Forward Optimization No. 3. Maximum consecutive loss equal to 83 losses in a row.}
   \label{fig: 1}
\end{figure}

\begin{figure*}[ht!]
  \centering
  \captionsetup{justification= justified}  
  \includegraphics[width=\textwidth, height = 11cm]{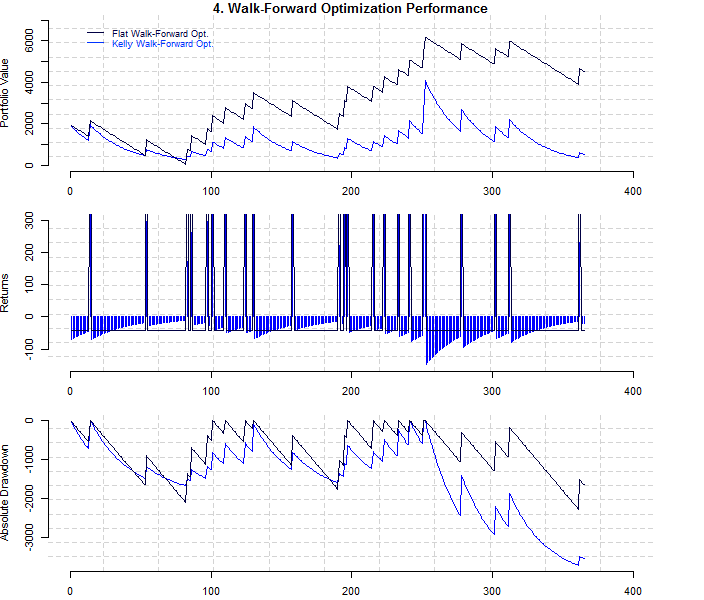}
    \caption{ Wagers on numbers 9 and 10. Probabilities of occurrence equal to $ 3.1961\%$ and $ 3.0546\%$ respectively. FBS' position size is $ \$42.49 $. Kelly's position size is $ 3.5722\% $. Probabilities gathered from in-sample analysis of 8,479 observations}
\end{figure*}

\begin{table}[H]
\centering
\label{my-label}
\caption{Performance Summary}
\begin{tabular}{c c c}
\hline
\textit{Statistics} & \textit{Kelly Criterion} & \textit{Flat wager}\\ [0.5ex]
\hline \\ 
Number of obs. &  365		&	365			\\
\\
Wins           & 23			&   23                   \\
\\
Losses         & 342		&   342                   \\
\\
P\&L           & -1,459.41	&   2,528.17           \\
\\
\% P\&L        & -72.97     &   126.41            \\
\\
Total Wins     & 13,330.28  &   17,102.31               \\
\\
Total Losses   & 14,718.25  &   14,531.65            \\
\\
Avg. Win       & 579.58     &   743.58               \\
\\
Avg. Loss      & 43.04      &   42.49              \\
\\
Max. Win       & 1572.00      &  743.58                 \\
\\
Max. Loss      & 145.90      &   42.49              \\
\\
Max. Drawdown  & 3,714.64   &   2,273.23              \\
\\
Calmar Ratio   & -0.39      &   1.11           		\\ \hline 
\end{tabular}
\end{table}

\begin{figure}[H]
  \centering
  \captionsetup{justification= justified}
  \includegraphics[width=3.5in, height = 4in]{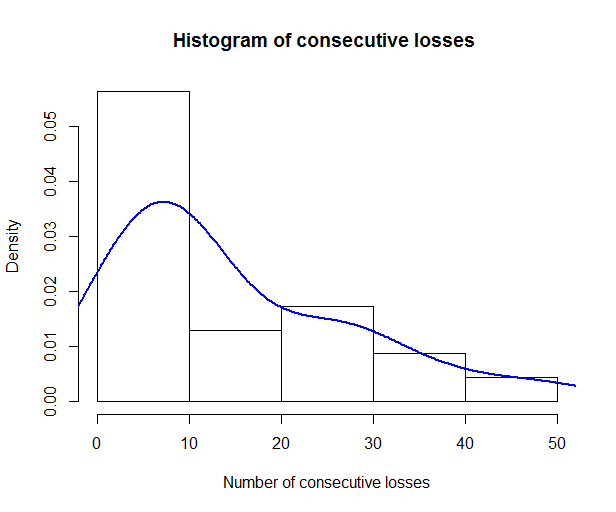}
    \caption{Histogram of consecutive losses for Walk-Forward Optimization No. 4. Maximum consecutive loss equal to 49 losses in a row.}
   \label{fig: 1}
\end{figure}

\begin{figure*}[ht!]
  \centering
  \captionsetup{justification= justified}  
  \includegraphics[width=\textwidth, height = 11cm]{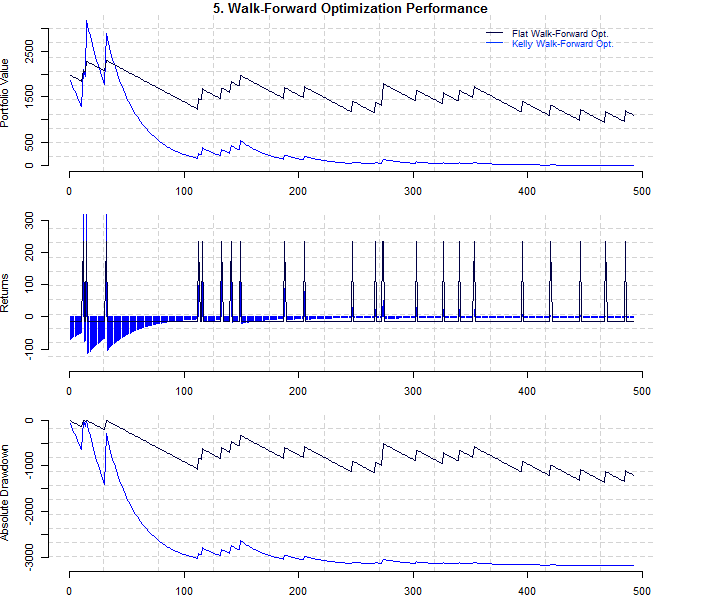}
    \caption{ Wagers on numbers 9 and 10. Probabilities of occurrence equal to $ 3.1547\%$ and $ 3.0981\%$ respectively. FBS' position size is $ \$13.33 $. Kelly's position size is $ 3.5743\% $. Probabilities gathered from in-sample analysis of 8,844 observations}
\end{figure*}

\begin{table}[H]
\centering
\label{my-label}
\caption{Performance Summary}
\begin{tabular}{c c c}
\hline
\textit{Statistics} & \textit{Kelly Criterion} & \textit{Flat wager}\\ [0.5ex]
\hline \\ 
Number of obs. &  492		&	492			\\
\\
Wins           & 23			&   23                   \\
\\
Losses         & 469		&   469                  \\
\\
P\&L           & -1,994.70	&   -899.51           \\
\\
\% P\&L        & -99.73     &   -44.98            \\
\\
Total Wins     & 4,327.74  &   5,263.73              \\
\\
Total Losses   & 6,250.95  &   6,249.91            \\
\\
Avg. Win       & 188.16     &   233.21              \\
\\
Avg. Loss      & 13.33      &   13.33              \\
\\
Max. Win       & 1,222.00      &  233.21                 \\
\\
Max. Loss      & 113.50      &   13.33              \\
\\
Max. Drawdown  & 3,170.51  &   1,359.26              \\
\\
Calmar Ratio   & -0.63      &   -0.66           		\\ \hline 
\end{tabular}
\end{table}

\begin{figure}[H]
  \centering
  \captionsetup{justification= justified}
  \includegraphics[width=3.5in, height = 4in]{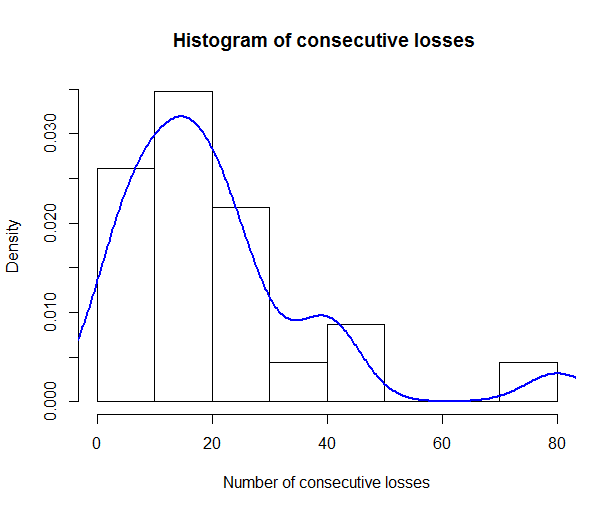}
    \caption{Histogram of consecutive losses for Walk-Forward Optimization No. 5. Maximum consecutive loss equal to 80 losses in a row.}
   \label{fig: 1}
\end{figure}

\begin{figure*}[ht!]
  \centering
  \captionsetup{justification= justified}  
  \includegraphics[width=\textwidth, height = 11cm]{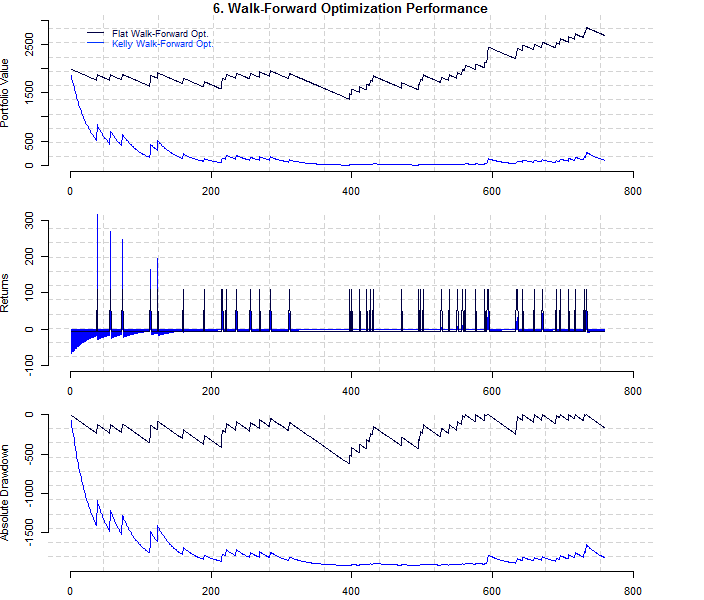}
    \caption{ Wagers on numbers 9 and 10. Probabilities of occurrence equal to $ 3.1170\%$ and $ 3.0527\%$ respectively. FBS' position size is $ \$6.27 $. Kelly's position size is $ 3.4888\% $. Probabilities gathered from in-sample analysis of 9,336 observations}
\end{figure*}

\begin{table}[H]
\centering
\label{my-label}
\caption{Performance Summary}
\begin{tabular}{c c c}
\hline
\textit{Statistics} & \textit{Kelly Criterion} & \textit{Flat wager}\\ [0.5ex]
\hline \\ 
Number of obs. &  759		&	759			\\
\\
Wins           & 47			&   47                   \\
\\
Losses         & 712		&   712                  \\
\\
P\&L           & -1,892.20	&   686.10           \\
\\
\% P\&L        & -94.61     &   34.31           \\
\\
Total Wins     & 2,714.65  &   5,153.59             \\
\\
Total Losses   & 4,537.08  &   4,461.22            \\
\\
Avg. Win       & 57.76    &   109.65              \\
\\
Avg. Loss      & 6.37      &   6.26              \\
\\
Max. Win       & 316.70      &  109.65                 \\
\\
Max. Loss      & 67.34      &   6.26              \\
\\
Max. Drawdown  & 1,924.77  &   626.58             \\
\\
Calmar Ratio   & -0.98      &   1.10           		\\ \hline 
\end{tabular}
\end{table}

\begin{figure}[H]
  \centering
  \captionsetup{justification= justified}
  \includegraphics[width=3.5in, height = 4in]{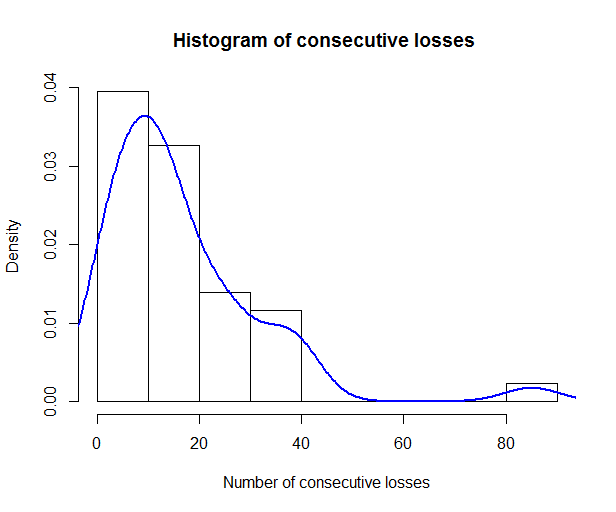}
    \caption{Histogram of consecutive losses for Walk-Forward Optimization No. 6. Maximum consecutive loss equal to 85 losses in a row.}
   \label{fig: 1}
\end{figure}

\begin{figure*}[ht!]
  \centering
  \captionsetup{justification= justified}  
  \includegraphics[width=\textwidth, height = 11cm]{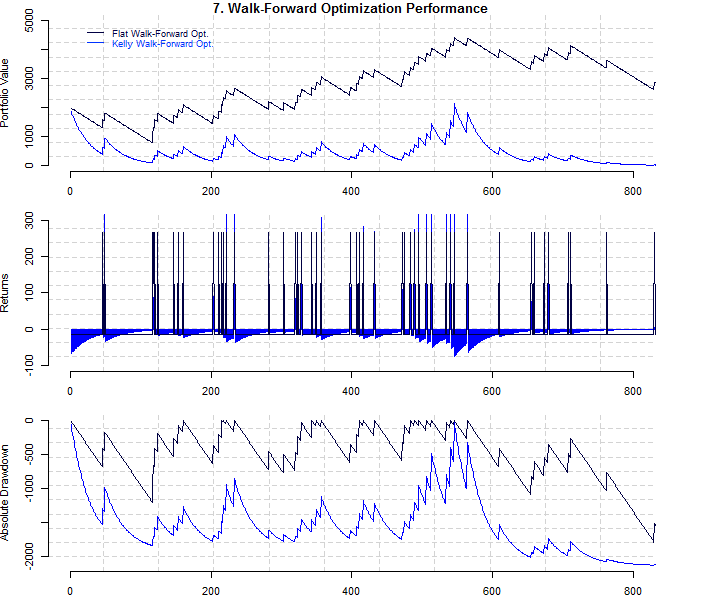}
    \caption{ Wagers on numbers 10 and 9. Probabilities of occurrence equal to $ 3.1105\%$ and $ 3.0609\%$ respectively. FBS' position size is $ \$15.25 $. Kelly's position size is $ 3.4906\% $. Probabilities gathered from in-sample analysis of 10,095 observations}
\end{figure*}

\begin{table}[H]
\centering
\label{my-label}
\caption{Performance Summary}
\begin{tabular}{c c c}
\hline
\textit{Statistics} & \textit{Kelly Criterion} & \textit{Flat wager}\\ [0.5ex]
\hline \\ 
Number of obs. &  885		&	885			\\
\\
Wins           & 50			&   50                  \\
\\
Losses         & 835		&   835                  \\
\\
P\&L           & -1,994.323	&   594.66           \\
\\
\% P\&L        & -99.72     &   29.73           \\
\\
Total Wins     & 10,944.32  &   13,341.75            \\
\\
Total Losses   & 12,868.84  &   -12,731.84            \\
\\
Avg. Win       & 218.89    &   266.83             \\
\\
Avg. Loss      & 15.41      &   15.25             \\
\\
Max. Win       & 811.80      &  266.83                \\
\\
Max. Loss      & 74.72      &   15.25             \\
\\
Max. Drawdown  & 2,135.71  &   1913.59             \\
\\
Calmar Ratio   & -0.93      &   0.31           		\\ \hline 
\end{tabular}
\end{table}

\begin{figure}[H]
  \centering
  \captionsetup{justification= justified}
  \includegraphics[width=3.5in, height = 4in]{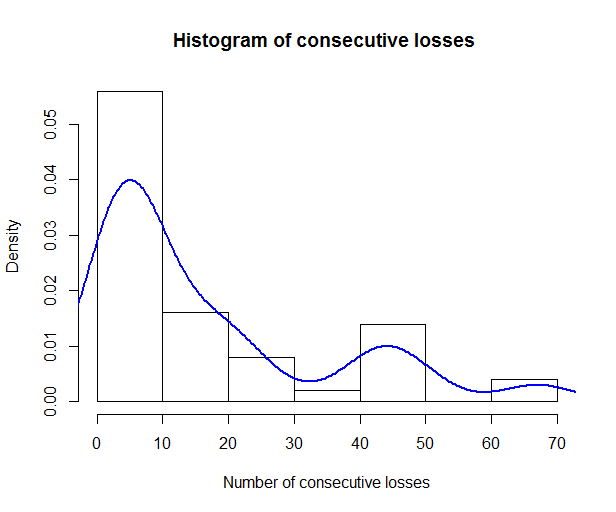}
    \caption{Histogram of consecutive losses for Walk-Forward Optimization No. 7. Maximum consecutive loss equal to 67 losses in a row.}
   \label{}
\end{figure} 

\clearpage
\clearpage

\section{Conclusion}

In this paper, we have seen that is not necessary that the roulette wheel has to be statistically biased in order to have profits in the short-term. Importantly, however, it is needed a probability of occurrence approximately to greater or equal than three percent for make a bet on a specific number. \\

Moreover, the probability of occurrence of each number at the roulette wheel can be modelled very accurately as an Ornstein-Uhlenbeck process. This result showed consistency with the Monte Carlo simulations shown in Figure 3. Further analysis is needed in order to assess if it is possible to make forecasts of how this probabilities evolve over time with this same model. \\

Furthermore, the flat betting system (FBS) has shown a better performance than the Kelly criterion's system, due that Kelly's carries higher volatility and it is more aggressive than the FBS, thus, as more losses in a row comes into the strategy it is more difficult for Kelly's system to recover. It would be a matter of further study to assess the performance of \textit{half} a Kelly criterion as a position size. Similarly, another matter of further study might involve perform a \textit{Non-anchored} Walk-forward optimization, in order to get more \textit{up-to-date} probabilities of occurrence and include the \textit{momentum} factor of the roulette wheel into the strategy. \\

In addition, it has also been observed that roulette outcomes get less predictable over time, which may be an evidence of the Second Law of Thermodynamics, which states that disorder, known as \textit{entropy}, increases with time. 

\section{Acknowledgements}

This research paper would not have been possible without the support of the following people, whom I am grateful: \\

\noindent My family, in alphabetical order: my brother, father and mother. For their tremendous support in my personal life over time. \\

\noindent My ex-wife, for her big support during the five years of my undergraduate education and sharing those years of her life with me. \\ 

\noindent My three godsons, in alphabetical order: Anthony, C\'esar and Piero, for took the crazy decision of choo\-sing me as their godfather. \\

\noindent Pedro Cornejo, mentor and friend who brought me into the financial industry, to whom I am deeply indebted. \\

\noindent Gabriel Chirre, for his loyal and disinterested friendship during my professional career. \\

\noindent David Hafermann, who taught me the concepts of backtesting and walk-forward optimization, and showed me the importance of programming. \\

\noindent Dennis La Cotera, a true Peruvian quant, who taught me the amazing properties of computational finance and applied mathematics in Wolfram Mathematica. \\

\noindent My professors at the CQF Institute, for sharing their great knowledge about quantitative finance with me. \\

\noindent My professors at Universidad San Ignacio de Loyola, who gave me the very first concepts of finance and econometrics at the Faculty of Economics. \\

\noindent In order of meeting to: Megija, Darja, Laura, Sigita and Jana, blackjack and roulette dealers at Evolution Gaming, for their invaluable motivational words that made this research paper possible. In this group, a very special thanks to Jana Zidere for verify in a daily basis my progress on this paper and for share the same crazy spirit as the author.   \\

\noindent Above all, thanks to the \textit{Intelligent Designer}.

\section{References}
[1] Pozdnyakov, Vladimir and Steele, Michael (2009). Martingale Methods for Patterns and Scan Statistics. \textit{Statistics for Industry and Technology}. pp. 289-317 \\

\noindent [2] Farmer, Doyne and Sidorowich John (1987). Predicting Chaotic Time Series. \textit{Physical Review Letters}. Vol 59, No. 8. \\

\noindent [3] Small, Michael and Tse, Chi Kong (2012). Predicting the outcome of roulette. \textit{Chaos: An Interdisciplinary Journal of Nonlinear Science} \\

\noindent [4] Jirina, Miroslav (1978). A biased roulette. \textit{Annales de l'Institute Henri Poincar\'e}. Vol 14, No 1, p. 1-23. \\

\noindent [5] Sundali, James and Croson, Rachel (2006). Biases in casino betting: The hot hand and the gambler's fallacy. \textit{Judgement and Decision Making}. Vol. 1, No. 1, pp. 1-12. \\

\noindent [6] Ethier, S.N. (1982). Testing for favourable numbers on a roulette wheel. \textit{Journal of the American Statistical Association}. Vol. 77, No. 379, pp. 660-665. \\

\noindent [7] Etemadi, N. Z. (1981). An elementary proof of the strong law of large numbers. \textit{Zeitschrift f\"ur Wahrscheinlichkeitstheorie und Verwandte Gebiete}. Vol. 55, Issue 1, pp. 119-122. \\

\noindent [8] Ni, Jiarui and Zhang, Chengqi (2005). An Efficient Implementation of the Backtesting of Trading Strategies. \textit{Lecture Notes in Computer Science}. Vol. 3758, pp. 126-131. \\ 

\noindent [9] Dixon M., Klabjan, D. and Bang, J. (2015). Backtesting Trading Strategies with Deep Neural Networks on the Intel Xeon Phi. \textit{SSRN Electronic Journal}. \\

\noindent [10] Wong. W (2008). Backtesting trading risk of commercial banks using expected shortfall. \textit{Journal of Banking and Finance}. Vol, 32, Issue 7, pp. 1404-1415. \\

\noindent [11] Menkveld, Albert J. (2013). High frequency trading and the new market makers. Vol. 16, Issue 4, pp. 712-740. \\

\noindent [12] Schneider, D. (2011). Trading at the speed of light. \textit{IEEE Spectrum}. Vol. 48, Issue 10, pp. 11-12. \\

\noindent [13] Scholtus M., and Dijk D. (2012). High-Frequency Technical Trading: The Importance of Speed. \textit{Tinbergen Institute Discussion Paper}. p. 63. \\

\noindent [14] Kelly, John L. Jr. (1956). A New Interpretation of Information Rate. \textit{Bell Systems Technical Journal}. Vol. 35, pp. 917-926. \\

\noindent [15] Young, T. (1991). Calmar Ratio: A Smoother Tool. \textit{Future Magazine}. Vol. 20, Issue 1, p. 40. \\

\noindent [16] Barndorff-Nielsen, O. and Shephard, N. (2001). Non-Gaussian Ornstein-Uhlenbeck-Based models and some of their uses in financial economics. \textit{Journal of the Royal Statistical Society. Series B (Statistical Methodology)} Vol. 63, Issue 2, pp. 167-241. \\

\noindent [17]  Maller R., M\"uller G. and Szimayer A. (2009). Ornstein-Uhlenbeck Processes and Extensions. \textit{Handbook of Financial Time Series}. Issue 1930, pp. 421-437.   \\

\noindent [18] \"Onalan, \"O. (2009). Financial Modelling with Ornstein-Uhlenbeck Processes Driven by L\'evy Process. \textit{Lecture Notes in Engineering and Computer Science}. Vol. 2177, Issue 1, pp. 1350-1355. \\

\end{document}